\documentclass{ws-ijmpd}
\usepackage[super,compress]{cite}
\usepackage{braket}
\begin{document}

\markboth{Asim Ullah, Jameel-Un Nabi and Muhammad Riaz}
{Effect of Nuclear Deformation on Electron Capture Cross-section on Chromium Isotopes}

%
\catchline{}{}{}{}{}
%

\title{Effect of Nuclear Deformation on Electron Capture Cross-section on Chromium Isotopes}

\author{Asim Ullah\footnote{corresponding author email:asimullah844@gmail.com }, Jameel-Un Nabi and Muhammad Riaz }
\address{Faculty of Engineering Sciences, Ghulam Ishaq Khan Institute of Engineering Sciences and Technology, Topi-23640, KP, Pakistan.\\
}

\maketitle

\begin{history}
\received{Day Month Year}
\revised{Day Month Year}
\end{history}

\begin{abstract}
The electron capture plays significant role in the pre-supernova and supernova evolutions of massive stars which in turn are of great importance in synthesizing heavy elements beyond iron. In this paper we study the effect of nuclear deformation  on the computed electron capture cross-section on selected even-even chromium isotopes ($^{46,48,50}$Cr). The nuclear deformation parameters were computed using two different theoretical models: Interacting Boson Model (IBM-1) and Macroscopic (Yukawa-plus-exponential) - microscopic (Folded-Yukawa) model (Mac-mic model). A third value of deformation parameter was adopted from experimental data. We chose the pn-QRPA model to perform our calculations. The predictive power of the chosen model was first tested by calculating Gamow-Teller (GT) strength distributions of selected $fp$-shell nuclei  where measured GT data was available. The calculated GT strength distributions were well fragmented over the energy range 0-12 $MeV$ and were noted to be in decent agreement with experimental data.
The total GT strength was found to increase (decrease) with decrease (increase) in the value of deformation parameter for the three chromium isotopes. The computed GT strength distributions satisfied the model independent Ikeda sum rule. The ECC were calculated as a function of the deformation parameter at core temperature 1.0 $MeV$. Our results show that the calculated ECC increased with increasing value of nuclear deformation. 

\end{abstract}




\section{Introduction}
The phenomenon in which stars end their lives is called supernova.
Study of supernova process is one of the means to explore our Universe. All type of natural interactions manifest themselves in a supernova explosion. 
Evolution of heavy mass stars and the supernova explosion are linked with the formation of heavy elements\cite{G.M.Fuller1980,E.M.Bur1957}. 
It is desirable that the computation of neutrino-induced reactions, stellar nucleosynthesis and supernova explosions be performed using microscopic global predictions instead of semi-empirical techniques. 
The electron capture cross-section (ECC) on iron group nuclei and electron capture (EC) on free protons play a crucial part in the pre-collapse  stages of massive stars.
At low temperatures ($300-800$~keV) and densities ($\sim 10^{10}~g/cm^3$) the Gamow-Teller (GT) strength is sensitive to EC rates and under such physical conditions the chemical potential and binding energy have comparable magnitudes. In this low temperature region the EC occurs on nuclei having $A \leq 60$. For high temperature and density regions EC occurs on nuclei having $A \geq 65$ and depends more on total GT strength  of the charge-changing transitions. 
Therefore the computation of GT strength distributions and ECC rates in stellar matter is a crucial requirement.\\
Recently the proton-neutron quasi-particle random phase approximation (pn-QRPA) model was used to compute ECC in stellar matter for even–even, odd-odd and odd–A $fp$-shell nuclei~\cite{Nab19}. The stellar EC rates on $fp$-shell nuclei computed using pn-QRPA were, in general, noted to be bigger than former calculations at high stellar temperatures.

In this paper we  explore the role of nuclear deformation parameter ($\beta_2$) on the computed ECC using the pn-QRPA model on even-even chromium isotopes ($^{46, 48, 50}$Cr). We first check the validity of chosen model by calculating GT strength distributions of iron-regime nuclei ($^{42}$Ti, $^{46}$Cr, $^{50}$Fe and $^{54}$Ni) and comparing the results against measured data and previous calculations.

The next section discusses the necessary theoretical framework to perform our calculation. We present our result in Section~3. Here we first show the performance of the chosen model and later investigate the effect of changing $\beta_2$ parameter on pn-QRPA calculated ECC on isotopes of chromium. Conclusions are stated in Section~4.

\section{Formalism}

The following Hamiltonian was considered for calculating GT strength distributions within the pn-QRPA formalism
\begin{equation}
H^{QRPA} = H^{sp} + V^{pp}_{GT} + V^{ph}_{GT} + V^{pair},
\end{equation}
where $H^{sp}$, $V_{GT}^{pp}$, $V_{GT}^{ph}$ and $V^{pair}$ represent Hamiltonian of single particle, particle-particle GT force, particle-hole GT force and pairing potential, respectively. 
For the calculation of wave functions and energies of the single particle, the Nilson model with deformed basis was used \cite{nilson1955}. The pairing correlations among the nucleons were considered in the BCS approximation.  The GT interaction strength parameters were fitted such that the measured energy of the GT giant resonance was reproduced wherever available.  The computed GT strength distributions satisfied the model independent Ikeda sum rule\cite{Ikeda1963}. The oscillator constant for both neutrons and protons was taken as $\hbar\omega=41A^{1/3}$ while the Nilson-potential parameter was taken from \cite{ragnarson1984}. The traditional value of pairing gaps  $\vartriangle_p=\vartriangle_n={12/\sqrt A}$  MeV ~\cite{hardy09} was considered in this work. $Q$-values were taken from Ref. \cite{audi2017}. For further details and solution of "Eq. (1)" we refer to \cite{Nab19}.\\

The nuclear reduced transition probability  from parent state "$m$" to daughter state "$n$"  is given by
\begin{equation}\label{5}
B_{mn}=(g_A/g_V )^2 B(GT)_{mn} + B(F)_{mn}. 
\end{equation}
The values of $g_A/g_V$ and $D$ were taken as -1.254~\cite{hardy09} and 6143$s$ \cite{Nak10}, respectively. The terms $B(F)_{mn}$ and $B(GT)_{mn}$ in~``Eq.~(\ref{5})" stands for the Fermi and GT reduced transition probabilities, respectively and are expressed as
\begin{equation}
B(F)_{mn}=\dfrac{1}{2J_m+1} |\langle n\|\sum_it_+^i \|m\rangle|^2
\end{equation}
\begin{equation}
B(GT)_{mn}=\dfrac{1}{2J_m+1} |\langle n\|\sum_it_+^i\sigma^{\rightarrow i} \|m\rangle|^2,
\end{equation}
where $J_m$ and $\sigma^{\rightarrow i}$ shows the total spin of the parent state $|m \rangle $ and Pauli spin matrices, respectively while $t_+^i$ represents the iso-spin raising operator. The summation was taken over all nucleons inside the nucleus.

The weak-interaction Hamiltonian for the calculation  of electron capture is given by
\begin{equation}
\widehat{H}_{\omega}=\dfrac{G}{\sqrt2}j_\mu^{lept}\widehat{J}^{\mu},
\end{equation}
where G=$G_Fcos\theta_c$. $G_F$ and $\theta_c$  represent Fermi coupling constant and Cabibbo angle, respectively. The leptonic current $j_\mu^{lept}$ and hadronic current $\widehat{J}^{\mu}$ are given by
\begin{equation}
j_{\mu}^{lep}=\bar{\psi}_{\upsilon_{e}}(x)\gamma_{\mu}(1-\gamma_5)\psi_{\upsilon_{e}}(x)
\end{equation}
\begin{equation}
\hat{j}^{\mu}=\bar{\psi}_{p}(x)\gamma_{\mu}(1-c_{A}\gamma_5)\psi_{n}(x),
\end{equation}
where the $\psi_{\nu_e}(x)$ are the neutrino and anti-neutrino spinors.  
The calculation of ECC is based on the matrix elements between parent $\Ket{m}$ and daughter  $\ket{n}$ states.
\begin{equation}
\bra{n}|\widehat{H}_{\omega}|\ket{m}=\dfrac{G}{\sqrt2}l^{\mu}\int
d^3xe^{-i{q.x}}\bra{n}\widehat{J_{\mu}}\ket{m},
\end{equation}
The terms $q$ and $l^\mu e^{-iq.x}$ in the above equation represent the three-momentum transfer and leptonic matrix element, respectively. The latter were used for computing the transition matrix elements \cite{NPaar2009, walecka2004}. 
In this work we assumed the low momentum transfer approximation $q \longrightarrow 0$. Then the GT operator ($GT^+ =\sum_{m}\tau_m^+\sigma_m$) contributes dominantly to the total stellar ECC~\cite{walecka2004}. The total ECC, as a function of incident energy ($E_e$) of the projectile electron and stellar temperature ($T$), may then be calculated using
\begin{equation}\label{9}
\begin{split}
\sigma(E_e,T)=\dfrac{G_F^{2}cos^2\theta_c}{2\pi}\sum\limits_{m}F(E_e,Z)\dfrac{(2J_{m}+1)\exp{(-E_m/kT)}}{G(Z,A,T)}\\
\times
\sum\limits_{J,f}(E_e+E_m-E_n-Q)^{2}\dfrac{|\bra{m} GT^{+} \ket{n}|^2}{(2J_{m}+1)},
\end{split}
\end{equation}
where $F(Z,E_e)$ and $G(A,Z,T)$ stand for the well-known Fermi and nuclear
partition function, respectively.  The last term of~``Eq.~(\ref{9})" is the nuclear matrix element between final and initial states. 
The Fermi function $F(Z,E_e)$ was calculated using the prescription of Gove and Martin~\cite{gov71}.
The nuclear
partition functions were calculated using a recent recipe introduced by Refs. \cite{nabo2016,nabi2016}. 

The motivation of current work was to investigate the effect of deformation parameter on calculated ECC. Three different nuclear deformation parameters were employed in the current investigation. Two values of $\beta_2$ were calculated using the IBM-1~\cite{arima1976} model and Macroscopic (Yukawa-plus-exponential) - microscopic (Folded-Yukawa) model (Mac-mic model)~\cite{muller81}. The third value was adopted from experimental data. In the latter case the deformation parameter was determined using adopted values of the reduced electric quadrupole transition probability $B(E_2)\uparrow$\cite{raman2001}. The Mac-mic model calculated the electric quadrupole moment ($Q_2$). This value was later used to determine the  values of deformation parameter using the following relation
\begin{equation}
\beta_2=\frac{125\hspace{0.01in} Q_2}{1.44 \hspace{0.01in} Z \hspace{0.01in}A^{2/3}}.	
\end{equation}
The IBM-1 and Ref.\cite{raman2001} used the following equation for calculation of deformation parameter
\begin{equation}
\beta_2=\frac{4\pi}{3ZR_0^2} [\dfrac{B(E_2)\uparrow}{e^2} ]^{1/2},
\end{equation}
where $R_0^2=0.0144 A^{2/3}b$ and $B(E_2)\uparrow$ is in units of $e^2 b^2$. The  $B(E_2)\uparrow$ values were calculated using the IBM-1 model. The experimental values of  $\beta_2$ were taken from  Ref.\cite{raman2001} and is denoted by $\beta$(E$_2$) in this paper.

\section{Results and Discussion}

As discussed earlier we first decided to test the predictive power of the current pn-QRPA model. For this we chose four key iron-regime nuclei, $^{42}$Ti, $^{46}$Cr, $^{50}$Fe and $^{54}$Ni, and calculated the GT strength distributions of these nuclei. The clacuated distributions were compared with measured data and previous calculations.

\begin{figure}[pb]
	\centerline{\psfig{file=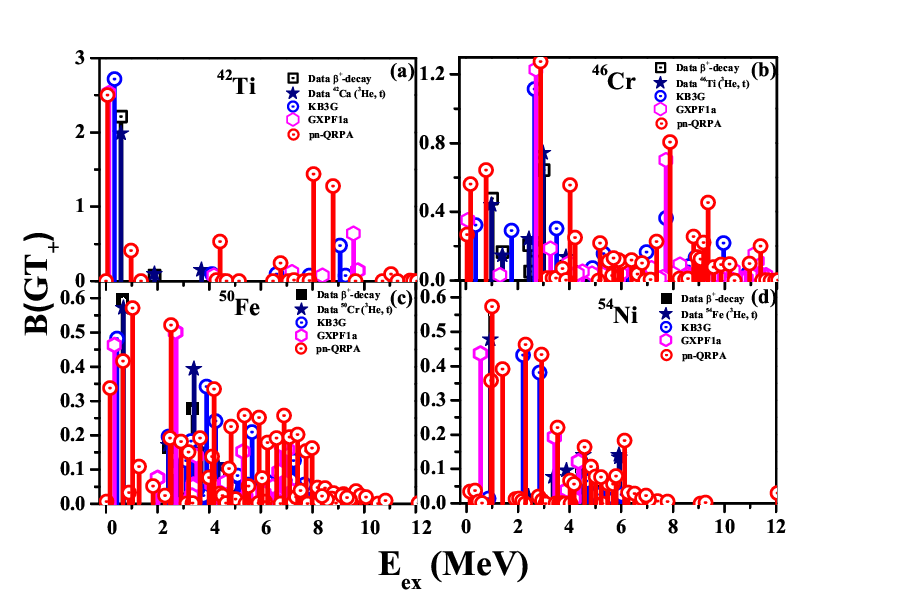,width=5.0in}}
	\vspace*{8pt}
	\caption{Calculated and measured GT strength distributions for $^{46}$Cr, $^{54}$Ni, $^{50}$Fe and $^{42}$Ti as a function of daughter excitation energy. For explanation of legends see text. \label{fig1}}
\end{figure}

Fig.~\ref{fig1} shows the measured \cite{Mol15, Ada06, Fuj05, Ada07, Ada12}, previous \cite{Kum16} and current calculation of GT strength distributions for  $^{42}$Ti, $^{46}$Cr, $^{50}$Fe and $^{54}$Ni. Fig.~\ref{fig1}(a) display the measured and computed GT strength distributions for $^{42}$Ti. The data of $\beta$-decay experiment was measured up to excitation energy of E$_{ex}=$ 1.888 $MeV$ in daughter. The data of charge exchange reaction $^{42}$Ca($^3$He, t) was measured up to E$_{ex}=$ 3.688 $MeV$. KB3G and GXPF1a interactions were used by shell model to compute the GT strength up to E$_{ex}=$ 12 $MeV$. Similarly GT strength was computed by pn-QRPA up to E$_{ex}=$ 12 $MeV$. The comparison shows that measured and computed data have decent agreement with each other. The GT strength calculated by pn-QRPA is well fragmented over  energy range (0-12) $MeV$ and the pn-QRPA strength is bigger than other measured and computed data. 

The GT strength for $^{46}$Cr is shown in Fig.~\ref{fig1}(b). In this case the excitation energies are 3.867 $MeV$, 5.717 $MeV$, and 12 $MeV$ for $\beta$-decay experiment, $^{46}$Ti($^3$He,t) and shell model calculation, respectively. The pn-QRPA result shows good comparison with  measured and computed data and the total calculated GT strength is bigger than  shell model results. 

The comparison of GT strength distribution for $^{50}$Fe is presented in Fig.~\ref{fig1}(c).   The  GT strength distribution was measured up to 4.315 $MeV$ excitation energy by $\beta$-decay experiment  and  up to E$_{ex}=$5.545 $MeV$ by charge exchange reaction. The shell model using KB3G and GXPF1a interactions and the current pn-QRPA model calculated GT transitions up to E$_{ex}=$12 $MeV$ are also shown in the figure.  The comparison shows that all the computed strength are in good  agreement with the measured data and the GT strength is well fragmented for $^{50}$Fe in the energy range of (0-12) $MeV$. The pn-QRPA calculated total strength up to 12 $MeV$ is bigger than the shell model calculated strengths.

The measured GT strength distributions for $^{54}$Ni up to excitation energy 5.202 $MeV$ by $\beta$-decay and E$_{ex}=$5.917 $MeV$ by $^{54}$Fe($^3$He, t) are shown in Fig.~\ref{fig1}(d).  The theoretical predictions using the shell model and pn-QRPA model are also shown up to 12 $MeV$ in daughter. The strength distributions show that the pn-QRPA computed strength is bigger than previous computed results and matched well  with the experimental  data. The results validated the use of pn-QRPA model for calculation of ECC rates.

Table~\ref{ta1} displays the comparison of measured  and computed total GT strengths for the above mentioned $fp$-shell nuclei. The $\beta-$decay and charge exchange reaction (CER)  data ($^3$He, t) are presented in third and fourth columns of Table.~1, respectively.  Columns 5 and 6 show the shell model results employing the  KB3G, GXPF1a interactions. Column~7 display the  extreme single particle model (ESPM)  results. For details of the ESPM model and calculation of GT strength we refer to \cite{Kum16}. The difference between measured and calculated data is attributed to the cut-off energies described before. Our calculated GT strengths are bigger than the shell model results. 

\begin{figure}[pb]
	\centerline{\psfig{file=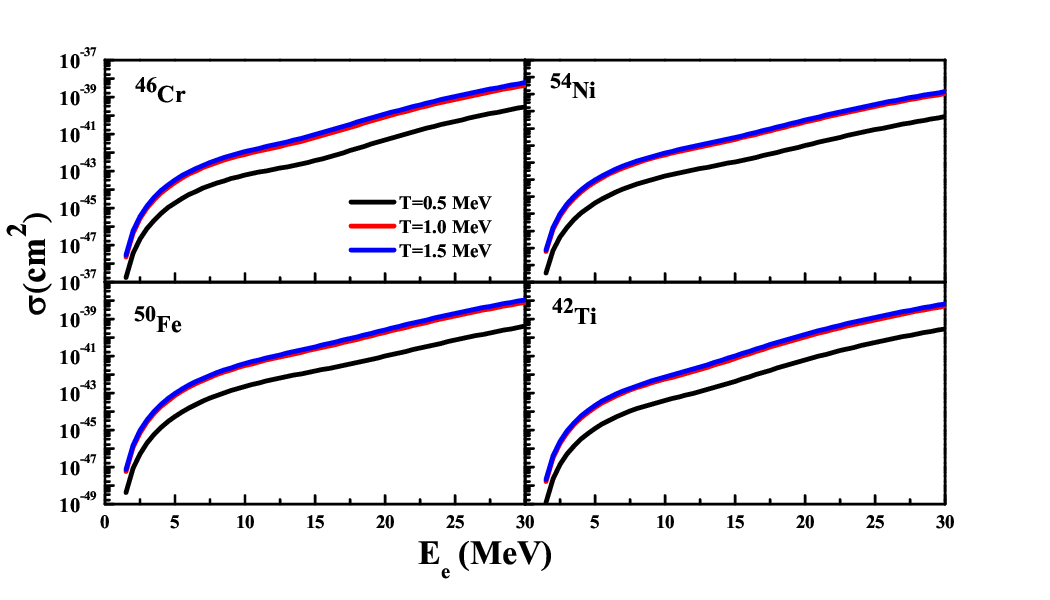,width=4.5in}}
	\vspace*{8pt}
	\caption{Calculated ECC, at three different temperature 0.5 $MeV$, 1 $MeV$ and 1.5 $MeV$, on $^{46}$Cr, $^{54}$Ni, $^{50}$Fe and $^{42}$Ti. \label{fig2}}
\end{figure}

We calculated ECC on selected $fp$-shell nuclei ($^{42}$Ti, $^{46}$Cr, $^{50}$Fe and $^{54}$Ni) employing the pn-QRPA model. The results are shown in Fig.~\ref{fig2}. The ECC for first few $MeV$ of incident electron energies increases very sharply. Later the increase becomes rather gradual. The trend may be explained by the calculated GT strength distributions presented earlier. The temperature effect on calculated ECC was also investigated. As the core temperature increases from 0.5 $MeV$ to 1.0 $MeV$, the computed ECC increased, on average,  by two orders of magnitude. This huge increment is attributed to the thermal unblocking of GT states. As the temperature further increases from 1.0 $MeV$ to 1.5 $MeV$ the calculated ECC increased marginally as the unblocking of states have already taken place. The trend of the calculated ECC is similar  for all four cases.
	 
Table~\ref{T1} displays the values of total strength and centroid of the pn-QRPA calculated GT strength distributions for the three isotopes of chromium. The pn-QRPA model nuclear deformation parameter, $\beta_2$,  was taken from Column~3 of Table~\ref{T1}. The computation of $\beta_2$ values was discussed in previous section. 
It is noted from Table~\ref{T1} that  the total GT  strength decreases for heavier isotopes of chromium. This is because as neutron number increases the EC process becomes more challenging. The effect of changing the deformation parameter can also be seen from the table. It is found that for each isotope the computed GT strength decreases with increase in the value of $\beta_2$. This finding, however, may not be generalized and further investigation is in progress.

\begin{figure}
	\centering
	\includegraphics[height=2.5in, width=3in]{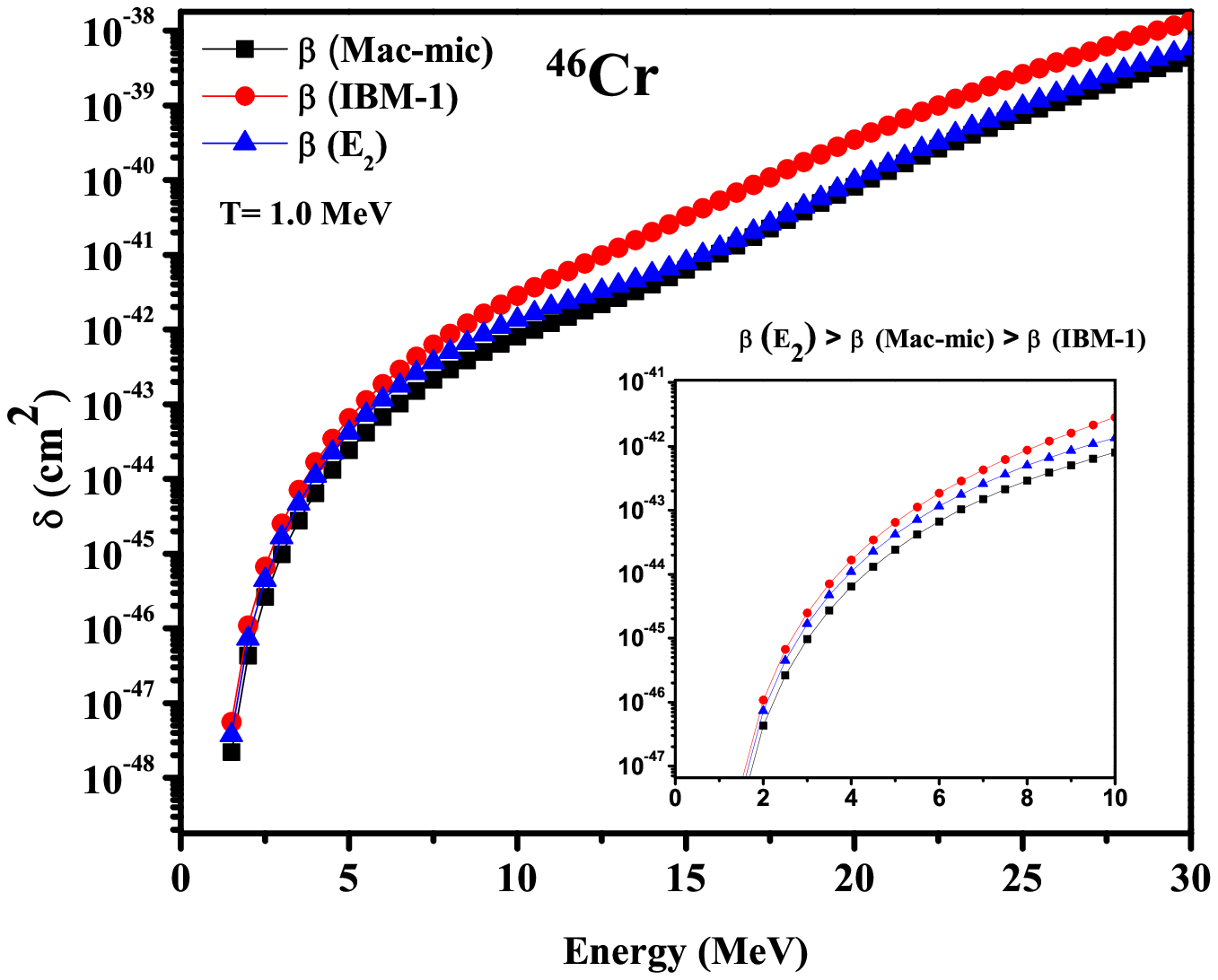}
	\caption{ Comparison of calculated ECC on $^{46}$Cr with different $\beta_2$ values. The inset shows the magnified data up to incident electron energy of 10 MeV.}
	\label{f4}
	\includegraphics[height=2.5in, width=3in]{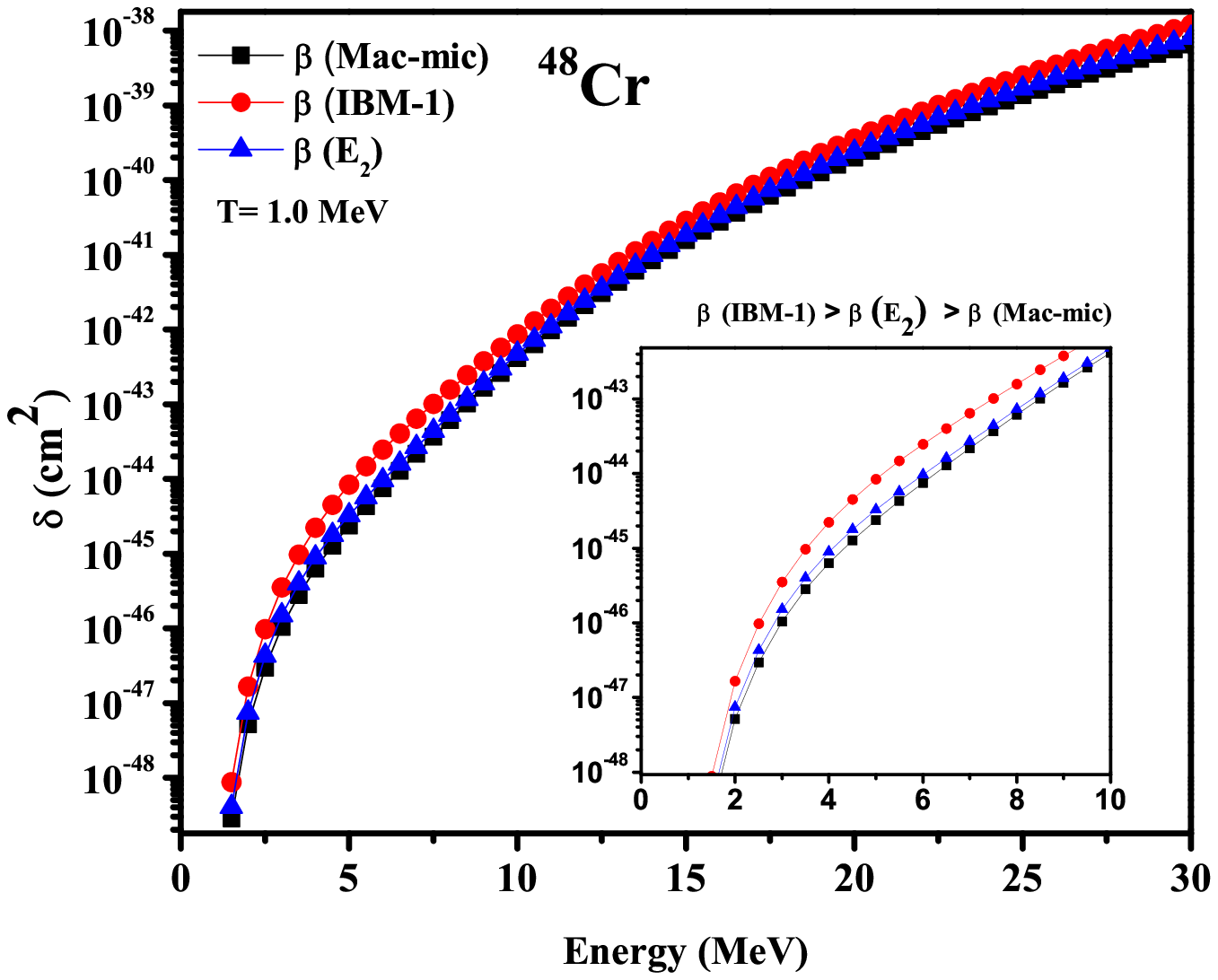}
	\caption{Same as Fig. 4, but for $^{48}$Cr.}
	\label{f5}
	\includegraphics[height=2.5in, width=3in]{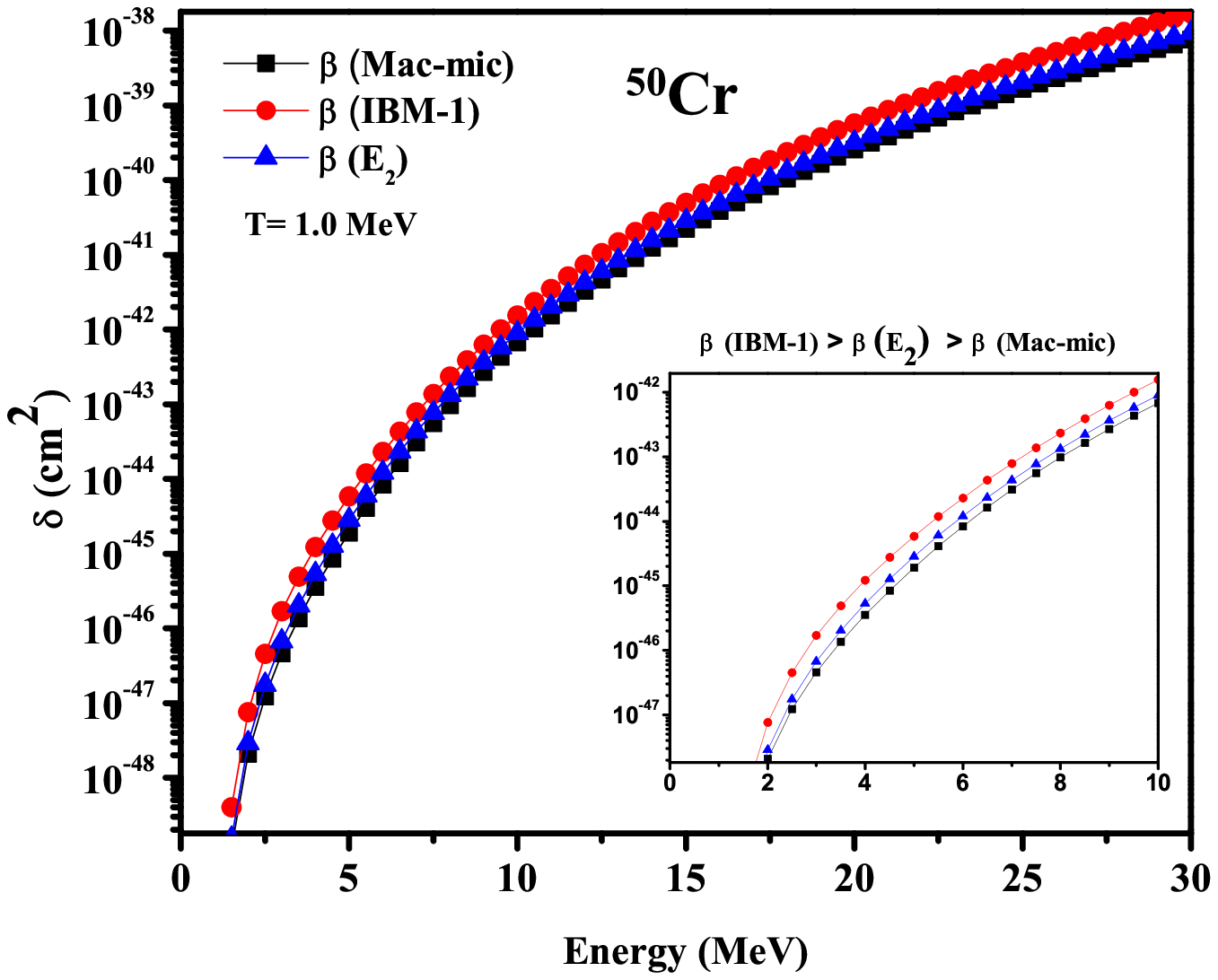}
	\caption{Same as Fig. 4, but for $^{50}$Cr.}
	\label{f6}
\end{figure}

Figs.~\ref{f4}-\ref{f6} investigate the effect of changing deformation parameter on calculated ECC on $^{46}$Cr, $^{48}$Cr and $^{50}$Cr, respectively. In Figs.~\ref{f4}-\ref{f6} we have fixed the stellar temperature to 1 MeV.  It is noted from these figures that with increase (decrease) in the magnitude of nuclear deformation of the three chromium isotopes, the ECC increases (decreases). We are not in a position to generalize this last statement which warrants further investigation. Effect of oblate deformation parameters on calculated ECC would also be explored in future.

\section{Conclusions}
The objective of current work was to investigate  effect of nuclear deformation on computed ECC. The required GT strength distributions were calculated using the pn-QRPA model. We first tested the predictive power of the pn-QRPA model. Four key iron-regime nuclei with measured GT strength were compared with the pn-QRPA model calculated GT strength distributions. We also compared pn-QRPA calculated distributions against previous calculations. Three even-even chromium isotopes ($^{46,48,50}$Cr) were selected and ECC at stellar temperatures T= (0.5 --1.5) MeV, employing the pn-QRPA model were computed. The calculated GT strength distributions satisfied the ISR and displayed a decent comparison with available experimental data. The GT strengths were fragmented and distributed well amongst the daughter states. The total GT strength and centroid in EC direction were calculated.  As the magnitude of nuclear deformation increased (decreased) the calculated total GT strength decreased (increased). The ECC was calculated at stellar temperature of (0.5, 1.0 and 1.5) MeV.  The calculated ECC increased with increase in temperature. The calculated ECC decreased (increased) with decrease (increase) in the values of deformation parameter. We are in a process of studying more cases (including oblate nuclei) and desire to report our  outcomes in near future. 
\section{Acknowledgment}
J.-U. Nabi would like to acknowledge the support of the
Higher Education Commission Pakistan through project
numbers 5557/KPK/NRPU/R$\&$D/HEC/2016 and 9-
5(Ph-1-MG-7)Pak-Turk/R$\&$D/HEC/2017 and Pakistan
Science Foundation through project number PSFTUBITAK/KP-GIKI (02).


\begin{table}[ph]
	\tbl{Comparison of measured and computed total GT strengths for selected $fp$-shell nuclei.}
	{\begin{tabular}{@{}cccccccc@{}} \toprule
			A & N & $\beta$-decay& CER& KB3G & GXPF1a& ESPM & pn-QRPA\\
			\colrule
			54    & 26    & 1.082 & 1.117 & 12.197 & 13.362& 16.29& 18.16  \\
			50    & 24     & 1.344 & 1.859 &  9.464 & 10.277& 14.14& 16.97\\
			46    & 22     & 2.047 & 2.219 &  7.231 &  7.613& 10.70&  9.50\\
			42    & 20     & 2.372 & 2.297 &  6.000 &  6.000&  6.00&  7.86\\
			\botrule
		\end{tabular} \label{ta1}}
\end{table}

\begin{table}[!h]
	\tbl{The calculated total GT strength and centroid values 
		in the  EC direction for the selected Cr isotopes with different deformation values.}
	{\begin{tabular}{@{}ccccc@{}} \toprule
			Nuclei & Model & $\beta_2$ & $\sum B(GT_+)$ &$\bar{E}_+(MeV)$ 
			\\
			
			\colrule
			$^{46}$Cr & Mac-mic& 0.02800& 9.13& 6.70\\
			& IBM-1  & 0.00000  & 9.67 & 5.83\\
			&Exp& 0.28800 & 8.97& 6.82\\
	
			\colrule
			$^{48}$Cr & Mac-mic& 0.23551& 8.94& 6.36\\
			& IBM-1  & 0.76000  & 8.46& 7.24 \\
			&Exp & 0.36800 & 8.63& 6.90\\
			
			\colrule
			$^{50}$Cr & Mac-mic& 0.14124& 8.58& 6.14\\
			& IBM-1  & 0.61200 & 7.22& 6.55 \\
			&Exp& 0.29000 & 7.66& 6.18\\
			
			\botrule
		\end{tabular}\label{T1}}
	\end{table}



\label{CDF}

\clearpage

\end{document}